\documentclass[aps,pre,amsmath,lengthcheck,superscriptaddress]{revtex4-2}

\usepackage{graphicx}\graphicspath{ {figures/} }
\usepackage{hyperref}
\hypersetup{colorlinks,allcolors=blue,breaklinks}

\newcommand{\be}{\begin{equation}}
\newcommand{\ee}{\end{equation}}
\newcommand{\ba}{\begin{align}}
\newcommand{\ea}{\end{align}}
\newcommand{\bi}{\begin{itemize}}
\newcommand{\ei}{\end{itemize}}

\newcommand{\tr}[1]{\mathrm{tr}\left\{#1\right\}}

\newcommand{\la}{\left\langle}
\newcommand{\ra}{\right\rangle}
\newcommand{\pd}{\partial}

\newcommand{\co}[1]{\cos{\left(#1\right)}}
\newcommand{\si}[1]{\sin{\left(#1\right)}}

\newcommand{\bla}{bla\\bla\\bla\\bla\\bla}

\newcommand{\mc}[1]{\mathcal{#1}}

\begin{document}

\title{Adiabatic processes like isothermal processes}

\author{Pierre Naz\'e}
\email{pierre.naze@unesp.br}

\affiliation{\it Departamento de F\'isica, Instituto de Geoci\^encias e Ci\^encias Exatas, Universidade Estadual Paulista ``J\'ulio de Mesquita Filho'', 13506-900, Rio Claro, SP, Brazil}

\date{\today}

\begin{abstract}

The objective of this work is to show that adiabatic processes can be very similar to isothermal ones. First, we show that the criteria for the compatibility of linear-response theory with the Second Law of Thermodynamics for thermally isolated systems are the same as those for systems performing isothermal processes. Motivated by such a result, we explore the thermodynamic consequences of the time-average excess work, observing an unexpected existence of a well-defined relaxation time for thermally isolated systems that obeys the Second Law of Thermodynamics. This is justified by recognizing that such systems, in the usual sense, present random relaxation time, which can be ``averaged'' by taking the time average of the relaxation function. Such a procedure is very similar to what happens in isothermal processes, where a stochastic average must be done on the relaxation function to have a well-defined relaxation time. In the end, we analyze the Landau-Zener model from this new point of view, discussing the construction of slowly-varying processes from linear-response theory and observing negative entropy production rates for non-monotonic and rapid protocols. 

\end{abstract}

\maketitle

\section{Introduction}
\label{sec:intro}

One of the main characteristics of isothermal processes is the existence of a relaxation timescale of the system. It allows, for instance, in the context of finite-time and weak drivings \cite{bonanca2014,acconcia2015a,acconcia2015b,bonanca2016,deffner2018,naze2020,naze2022}, a natural expansion of the thermodynamic work of weak processes to slowly-varying ones \cite{mou1994,antonelli1997,bonanca2014} and creates a reference parameter that characterizes the emergence of negative entropy production rates under certain conditions \cite{naze2020,deffner2021}. On the other hand, thermally isolated systems, which execute adiabatic processes, do not present a well-defined relaxation timescale, except, at least, for chaotic \cite{falcioni1990,ngai1991,frahm1997,fishman2002,agam2013, faria2020} and quantum many-body systems \cite{zurek2005,cramer2008,langen2013,eisert2015,schiulaz2019,naze2022b}. This leads to a natural disagreement between the behaviors of the thermodynamic work of both processes, like the behavior of the optimal excess work for long switching times \cite{crooks2012,naze2022,soriani2022}. 

We present however in this work that adiabatic processes can be very similar to isothermal ones in some contexts. Proving at the beginning that the criteria used to justify the compatibility of linear-response theory with the Second Law of Thermodynamics for isothermal processes remain the same for thermally isolated ones, we propose, in the context of finite-time, weak and adiabatic driving, that the quantity we should observe to prove our point is not the average work, but its time-averaged quantity. 

To understand this point, consider that in isothermal processes, the averaged work does not capture the randomness inherent in the stochastic nature of the system, and, in practice, the system presents a random relaxation time. Indeed, to obtain a meaningful result in that sense, a stochastic average must be done in the averaged work. In typical scenarios of thermally isolated systems with oscillatory relaxation functions \cite{acconcia2015b,acconcia2015a,bonanca2016, myers2021}, the same randomness on the relaxation time can be considered too. Inspired by what happens in isothermal processes, we expect that an appropriate average in the average work will furnish a meaningful result. In this case, it will be the average in the switching time of the process.

One of the main consequences of taking this average is the appearance of an unexpected, well-defined relaxation time for the system. It will allow these specific thermally isolated systems of having the same mathematical property as those performing isothermal processes. In other words, adiabatic processes will be treated on the same foot as isothermal processes. This will be illustrated with the Landau-Zener model, where the isothermal processes properties, like a slowly-varying process expansion and the appearance of negative values to the entropy production rates for non-monotonic and rapid drivings, will be recovered in this new time-averaged work approach.

\section{Excess work in linear response theory}
\label{sec:lrt}

Consider a quantum system with a Hamiltonian $\mc{H}(\lambda(t))$, where $\lambda(t)$ is a time-dependent external parameter. Initially, this system is in contact with a heat bath of temperature $\beta\equiv {(k_B T)}^{-1}$, where $k_B$ is Boltzmann's constant. The system is then decoupled from the heat bath and, during a switching time $\tau$, the external parameter is changed from $\lambda_0$ to $\lambda_0+\delta\lambda$. The average work performed on the system during this process is
\be
W \equiv \int_0^\tau \la\pd_{\lambda}\mc{H}(t)\ra\dot{\lambda}(t)dt,
\label{eq:work}
\ee
where $\partial_\lambda$ is the partial derivative for $\lambda$ and the superscripted dot is the total time derivative. The generalized force $\la\pd_{\lambda}\mc{H}(t)\ra$ is calculated using the trace over the density matrix $\rho(t)$
\be
\la A(t)\ra =\tr{A\rho(t)}
\ee
where $A$ is some observable. The density matrix $\rho(t)$ evolves according to Liouville equation
\be
\dot{\rho} =\mc{L}\rho:= -\frac{1}{i\hbar}[\rho,\mc{H}],
\ee
where $\mc{L}$ is the Liouville operator, $[\cdot,\cdot]$ is the commutator and $\rho(0)=\rho_c$ is the initial canonical density matrix. Consider also that the external parameter can be expressed as
\be
\lambda(t) = \lambda_0+g(t)\delta\lambda,
\ee
where to satisfy the initial conditions of the external parameter, the protocol $g(t)$ must satisfy the following boundary conditions
\be
g(0)=0,\quad g(\tau)=1. 
\label{eq:bc}
\ee

Linear response theory aims to express the average of some observable until the first order of some perturbation considering how this perturbation affects the observable and the non-equilibrium density matrix \cite{kubo2012}. In our case, we consider that the parameter does not considerably changes during the process, $|g(t)\delta\lambda/\lambda_0|\ll 1$, for all $t\in[0,\tau]$. Using the framework of linear-response theory \cite{kubo2012}, the generalized force $\la\pd_{\lambda}\mc{H}(t)\ra$ can be approximated until the first-order as
\begin{equation}
\begin{split}
\la\pd_{\lambda}\mc{H}(t)\ra =&\, \la\pd_{\lambda}\mc{H}\ra_0+\delta\lambda\la\pd_{\lambda\lambda}^2\mc{H}\ra_0 g(t)\\
&-\delta\lambda\int_0^t \phi_0(t-t')g(t')dt',
\label{eq:genforce-resp}
\end{split}
\end{equation}
where the $\la\cdot\ra_0$ is the average over the initial canonical density matrix. The quantity $\phi_0(t)$ is the so-called response function \cite{kubo2012}, which can be conveniently expressed as the derivative of the relaxation function $\Psi_0(t)$ \cite{kubo2012}
\be
\phi_0(t) = -\frac{d \Psi_0}{dt},
\label{eq:resprelax}
\ee 
where
\be
\Psi_0(t)=\beta\langle\partial_\lambda\mathcal{H}(t)\partial_\lambda\mathcal{H}(0)\rangle_0+\mathcal{C},
\ee
being the constant $\mathcal{C}$ calculated via the final value theorem \cite{kubo2012}. In this manner, the generalized force, written in terms of the relaxation function, is
\begin{equation}
\begin{split}
\la\pd_{\lambda}\mc{H}(t)\ra =&\, \la\pd_{\lambda}\mc{H}\ra_0-\delta\lambda\widetilde{\Psi}_0 g(t)\\
&+\delta\lambda\int_0^t \Psi_0(t-t')\dot{g}(t')dt',
\label{eq:genforce-relax}
\end{split}
\end{equation}
where $\widetilde{\Psi}_0(t)\equiv \Psi_0(0)-\la\pd_{\lambda\lambda}^2\mc{H}\ra_0$. Combining Eqs. (\ref{eq:work}) and (\ref{eq:genforce-relax}), the average work performed at the linear response of the generalized force is
\begin{equation}
\begin{split}
W = &\, \delta\lambda\la\pd_{\lambda}\mc{H}\ra_0-\frac{\delta\lambda^2}{2}\widetilde{\Psi}_0\\
&+\delta\lambda^2 \int_0^\tau\int_0^t \Psi_0(t-t')\dot{g}(t')\dot{g}(t)dt'dt.
\label{eq:work2}
\end{split}
\end{equation}

We remark that in thermally isolated systems, the work is separated into two contributions: the quasistatic work $W_{\rm qs}$ and the excess work $W_{\rm ex}$. We observe that only the double integral on Eq.~(\ref{eq:work2}) has ``memory'' of the trajectory of $\lambda(t)$. Therefore the other terms are part of the contribution of the quasistatic work. Thus, we can split them as
\be
W_{\rm qs} = \delta\lambda\la\pd_{\lambda}\mc{H}\ra_0-\frac{\delta\lambda^2}{2}\widetilde{\Psi}_0,
\ee  
\begin{equation}
\begin{split}
W_{\text{ex}} = \delta\lambda^2 \int_0^\tau\int_0^t \Psi_0(t-t')\dot{g}(t')\dot{g}(t)dt'dt.
\label{eq:wirrder0}
\end{split}
\end{equation}
In particular, the excess work can be rewritten using the symmetry property of the relaxation function, $\Psi(t)=\Psi(-t)$ (see Ref.~\cite{kubo2012}),
\begin{equation}
\begin{split}
W_{\text{ex}} = \frac{\delta\lambda^2}{2} \int_0^\tau\int_0^\tau \Psi_0(t-t')\dot{g}(t')\dot{g}(t)dt'dt.
\label{eq:wirrder}
\end{split}
\end{equation}

We remark that such treatment can be applied to classic systems, by changing the operators to functions, and the commutator by the Poisson bracket \cite{kubo2012}.

\section{Isothermal processes}
\label{sec:isothermal}

The description of the previous section was made for adiabatic processes. However, the framework of linear-response theory can be applied similarly to isothermal processes, where a stochastic approach is appropriate. In this case, the average work $W$ is divided into the irreversible work $W_{\rm irr}$ and the difference of Helmholtz free energy $\Delta F$
\be
W = W_{\rm irr}-\Delta F,
\ee
where, in linear-response theory, we will have
\begin{equation}
\begin{split}
W_{\rm irr} = \frac{\delta\lambda^2}{2} \int_0^\tau\int_0^\tau \overline{\overline{\Psi}}_0(t-t')\dot{g}(t')\dot{g}(t)dt'dt,
\label{eq:wirrder}
\end{split}
\end{equation}
where $\overline{\overline{\cdot}}$ is the stochastic average. To satisfy the Second Law of Thermodynamics for isothermal processes, that is,
\be
\lim_{\tau\rightarrow\infty} W_{\rm irr}=0,\quad W_{\rm irr}(\tau)\ge 0, \forall \tau,
\ee
as it is explicitly deduced in Jarzynski's work \cite{jarzynski1997}, the relaxation function must satisfy the following criteria \cite{naze2020}
\be
\widetilde{\overline{\overline{\Psi}}}_0(0)<\infty,\quad \Hat{\overline{\overline{\Psi}}}_0(\omega)\ge 0,
\ee
where $\widetilde{\cdot}$ and $\Hat{\cdot}$ are respectively the Laplace and Fourier transforms. Because of this, it is possible to define a relaxation time for the system
\be
\tau_R := \int_0^\infty\frac{\overline{\overline{\Psi}}(t)}{\Psi(0)}dt = \frac{\widetilde{\overline{\overline{\Psi}}}_0(0)}{\Psi_0(0)}<\infty.
\ee

\begin{figure}
    \includegraphics[scale=0.4]{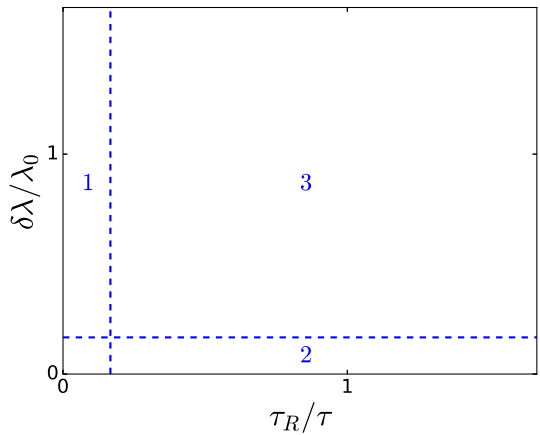}
    \caption{(Color online) Diagram of non-equilibrium regions. Region 1: slowly-varying processes, Region 2: finite-time but weak processes, and Region 3: arbitrarily far-from-equilibrium processes. Linear-response theory can describe regions 1 and 2.}
\label{fig:diagram}
\end{figure}

We discuss some consequences of the existence of such a quantity. First, it establishes a criterion to identify how fast the driving is performed. Indeed, one can create a diagram of non-equilibrium regions illustrating that. See Fig.~\ref{fig:diagram}. In region 1, the so-called slowly-varying processes, the ratio $\delta\lambda/\lambda_0$ is arbitrary, while $\tau_R/\tau\ll 1$. By contrast, in region 2, the so-called finite-time and weak processes, the ratio $\delta\lambda/\lambda_0\ll 1$, while $\tau_R/\tau$ is arbitrary. In region 3, the so-called arbitrarily far-from-equilibrium processes, both ratios are arbitrary. Linear-response theory can be used to calculate the irreversible work of regions 1 and 2.

Indeed, Eq.~\eqref{eq:wirrder} describes the irreversible work performed in region 2. In region 1, the approximation used to calculate the relaxation function is \cite{naze2022b}
\be
\lim_{\tau_R/\tau\ll 1}\overline{\overline{\Psi}}_0(t) = 2\tau_R \Psi_0(0)\delta(t),
\label{eq:approximationSL}
\ee
In this manner, the averaged irreversible work becomes \cite{bonanca2014}
\be
W_{\rm irr}(\tau) = \int_0^\tau \tau_R[\lambda(t)]\chi[\lambda(t)]\lambda(t)^2dt,
\ee
where 
\be
\chi[\lambda_0] = \Psi_0(0).
\ee

Second, the regions determined by the ratio of the relaxation time and switching time could present different behaviors when regarding the entropy production rates. Defining such quantity by
\be
\dot{W}_{\rm irr} = \dot{\lambda}(t)\int_0^t\overline{\overline{\Psi}}_0(t-t')\dot{\lambda}(t')dt',
\ee
we observe that its sign, for slowly-varying processes, is always positive, while for finite-time and weak processes, non-monotonic external protocols can produce instants of time where the sign is negative \cite{naze2020,deffner2021}. Even though, its integral along the switching time is always positive \cite{naze2020,deffner2021,bonanca2022}.

We remark that such characteristics raised in the previous paragraphs do not hold necessarily for thermally isolated systems, because the relaxation function does not decorrelate (see a typical example in Eq. \eqref{eq:randomrelaxtime}). In Sec.~\ref{sec:taew}, we present a new definition of work where those characteristics are recovered for a specific type of thermally isolated system.

\section{Criteria of compatibility for adiabatic processes}
\label{sec:isuues}

The first objective of this work is to find criteria for the compatibility of linear-response theory with the Second Law of Thermodynamics for thermally isolated systems. We validate in this manner the use of linear-response theory in describing thermodynamic processes in this context, as we have done in the context of isothermal processes \cite{naze2020}. In this way, one important aspect of Eq.~\eqref{eq:wirrder} is its resemblance with the expression of the irreversible work used to describe isothermal drivings in linear-response theory (see Eq. \eqref{eq:wirrder}). If the Second Law of Thermodynamics for thermally isolated systems can be stated as
\be
\lim_{\tau\rightarrow\infty}W_{\rm ex}(\tau)=0,\quad W_{\rm ex}(\tau)\ge 0, \forall \tau,
\label{eq:2ndstate}
\ee
as Jarzynski demonstrates in \cite{jarzynski2020}, then the relaxation function should have the same criteria described on Ref.~\cite{naze2020} to satisfy the above statements
\be
\widetilde{\Psi}_0(0)<\infty,\quad \Hat{\Psi}_0(\omega)\ge 0,
\label{eq:sufficient2ndlaw}
\ee
The second criterion is easily proven using Bochner's theorem \cite{naze2020,deffner2021}. However, the first criterion is based on the direct application of the final value theorem \cite{naze2020}. If its conditions are not satisfied, the compatibility with the vanishing behavior of the excess work for large times can not be proven by this criterion. Indeed, if we analyze a typical class of systems of thermally isolated systems, those presenting oscillatory relaxation function \cite{acconcia2015a,bonanca2016,myers2021},
\be
\Psi(t)=A\co{\omega t},
\ee
we observe that one of the poles of its Laplace transform is in the right complex plane, so the final value theorem cannot be applied.

However, there is an alternative version of the final value theorem that can be used in place of its traditional form \cite{gluskin2003}: If $\eta(t)$ is a bounded function, then
\be
\lim_{s\rightarrow 0^+}s\widetilde{\eta}(s)=\lim_{\tau\rightarrow\infty}\frac{1}{\tau}\int_0^\tau \eta(t)dt
\label{eq:appfvt}
\ee
Therefore, the finiteness of the Laplace transform can be used again to prove the asymptotic limit of Eq.~\eqref{eq:2ndstate}. Indeed 
\begin{align}
 \widetilde{\Psi}(0)<\infty &\Leftrightarrow \lim_{\tau\rightarrow\infty}\frac{1}{\tau}\int_0^\tau \Psi(t)dt=0\\
 &\Leftrightarrow \lim_{t\rightarrow \infty} \frac{1}{\tau}\int_0^\tau W_{\rm ex}(t)dt=0,
\end{align}
and
\begin{align}
\lim_{\tau\rightarrow \infty} \frac{1}{\tau}\int_0^\tau W_{\rm ex}(t)dt=0 &\Leftrightarrow \lim_{\tau\rightarrow \infty} W_{\rm ex}(\tau)=0.
\label{eq:equivasymplim}
\end{align}
where we consider the alternative version of the final value theorem. For more detail, see App.~\ref{app:A}. Therefore, the conditions \eqref{eq:sufficient2ndlaw} are enough to prove the compatibility of linear-response theory with the Second Law of Thermodynamics for thermally isolated systems with bounded relaxation functions.

\section{Time-averaged excess work}
\label{sec:taew}

The previous result highlights the role of the time-averaged excess work 
\be
\overline{W}(\tau)=\frac{1}{\tau}\int_0^\tau W(t)dt
\ee
to determine a criterion to make compatible the formalism of linear-response theory with the Second Law of Thermodynamics. In this manner, we investigate the thermodynamic consequences to adiabatic processes if we observe as the main quantity of interest the time-averaged thermodynamic work, instead of its traditional averaged form. Taking this new point-of-view, we shall see that this produces an unexpected relaxation time for the system. In this manner, in principle, an adiabatic process will have the same characteristics as an isothermal one. In Sec.~\ref{sec:example} we illustrate that presenting the two characteristics discussed in Sec.~\ref{sec:isothermal} with the Landau-Zener model. 

In the following sections, we present how time-averaged work can be calculated using linear-response theory and its relaxation time. To do so, we define the idea of time-averaged excess work
\be
\overline{W}_{\rm ex} = \frac{1}{\tau}\int_0^\tau W_{\rm ex}(t)dt,
\ee
where $W=W_{\rm ex}+W_{\rm qs}$.

\subsection{Equivalence between 2nd Laws}

First of all, it is important to see if the time-averaged work preserves the Second Law of Thermodynamics when observed independently from the averaged work. We are going to prove then that the properties of Eq.~\eqref{eq:2ndstate} are equivalent to
\be
\lim_{\tau\rightarrow\infty}\overline{W}_{\rm ex}(\tau)=0,\quad \overline{W}_{\rm ex}(\tau)\ge 0, \forall \tau.
\label{eq:2ndLawTimeAveragingExcessWork2}
\ee

The equivalence between Eqs.~\eqref{eq:2ndstate} and  \eqref{eq:2ndLawTimeAveragingExcessWork2} for the asymptotic limit was proven in Eq.~\eqref{eq:equivasymplim}. Concerning the equivalence of the positiveness of the excess works, we prove first the implication \eqref{eq:2ndstate}$\Rightarrow$\eqref{eq:2ndLawTimeAveragingExcessWork2}.  It is easy to see that
\be
\overline{W}_{\rm ex}(\tau) = \frac{1}{\tau}\int_0^\tau W_{\rm ex}(t) dt\ge 0.
\ee
Now we prove the second implication \eqref{eq:2ndLawTimeAveragingExcessWork2}$\Rightarrow$ \eqref{eq:2ndstate} of the positiveness of the excess works. Consider that there exists a particular switching time  $\tau_0$ where
\be
W_{\rm ex}(\tau_0)<0.
\ee
Since the excess work is a continuous quantity it must exist an interval $[\tau_0-\alpha,\tau_0+\gamma]$, with $\alpha>0$ and $\gamma>0$, where the excess work must be negative for all instants within it. It is then possible -- changing the initial time -- to determine a new time $\alpha+\gamma$ where $\overline{W}_{\rm ex}$ must be negative. Therefore, the implication is proved.

In this manner, we only have to regard the Eq.~\eqref{eq:2ndLawTimeAveragingExcessWork2} as a new statement of the 2nd Law of Thermodynamics.

\subsection{Linear-response theory}

Now we observe how the time-averaged excess work can be calculated using linear-response theory. One can easily show that (see App. \ref{app:B})
\be
\overline{W}_{\text{ex}}(\tau) = \frac{\delta\lambda^2}{2} \int_0^\tau\int_0^\tau \overline{\Psi}_0(t-t')\dot{g}(t')\dot{g}(t)dt'dt,
\label{eq:TAexcesswork}
\ee
where
\be
\overline{\Psi}_0(t)=\frac{1}{t}\int_0^t \Psi_0(u)du,
\label{eq:relaxfuncaverage}
\ee
is the time-averaged relaxation function. This means that calculating the time-averaged excess work is the same as calculating the averaged excess work, but with the time-averaged relaxation function.

\subsection{Time-averaged relaxation time}

When measured with time-averaged work, the system presents a relaxation time. Indeed, the conditions such that linear-response theory is compatible with the Second Law of Thermodynamics of Eq.~\eqref{eq:2ndLawTimeAveragingExcessWork2} are the same as those of Ref.~\cite{naze2020}
\be
\widetilde{\overline{\Psi}}_0(0)<\infty,\quad \Hat{\overline{\Psi}}_0(\omega)\ge 0.
\ee
Therefore, analogously to what happens in an isothermal process, we define a new relaxation time
\be
\overline{\tau}_R := \int_0^\infty\frac{\overline{\Psi}_0(t)}{\overline{\Psi}_0(0)}dt=\frac{\widetilde{\overline{\Psi}}_0(0)}{\overline{\Psi}_0(0)}<\infty.
\label{eq:averagedrelaxtime}
\ee

\subsection{Physical meaning}

Mathematically, we approximate thermally isolated systems to isothermal ones by giving the former a relaxation time. Physically speaking, what does that mean? In what follows we give a interpretation of what happens.

The isothermal process has as a main feature a stochastic process acting on the system of interest due to the dynamics of the heat bath. Because of this, the relaxation time of the system is ill-defined and presents random aspects. It is then necessary to take a stochastic average in every ensemble average performed on an observable of the system.  In particular, this stochastic average appears in the definition of the relaxation function of systems that passes through an isothermal process
\be
\overline{\overline{\Psi}}_0(t) = \beta \langle\partial_\lambda\mathcal{H}(0)\overline{\overline{\partial_\lambda \mathcal{H}}}(t)\rangle_0 - \mathcal{C}
\ee
If we observe the new relaxation function of the thermally isolated system defined in Eq.~\eqref{eq:relaxfuncaverage}, we have a new average on the generalized force, not in the stochastic sense, but in time. Is it possible that the thermally isolated system passes through a random process at each time, like in isothermal ones? We can not affirm this in general, but, as we will see in Sec.~\ref{sec:example}, the relaxation time of systems presenting oscillatory relaxation functions are mathematically ill-defined (see Eq.~\eqref{eq:randomrelaxtime}), and can be physically interpreted as a limited random number. Therefore, at each moment along the process, the system of interest presents a different relaxation time and, implicitly, a different relaxation function. In this manner, we have to average in time those relaxation functions to produce a new one which we expect will furnish a well-defined relaxation time to the system. Observing Eq.~\eqref{eq:averagedrelaxtime}, that is indeed the case. Also, as far as I know, there are no other situations of thermally isolated systems with well-defined relaxation times in the usual sense.

In the end, adiabatic and isothermal processes can be indeed very similar, when both have randomness in their relaxation time. In this case, this can be solved by taking appropriate averages in their relaxation functions, each one in its own way. In particular, the irreversible work can be seen as the stochastic average of the work given by Eq.~\eqref{eq:work}
\be
W_{\rm irr} = \overline{\overline{W}}-\Delta F.
\ee

\subsection{Executing in the laboratory}

In order to furnish a feasible way to calculate the time-averaged work without calculating the time-averaged relaxation function, we remark that the expression \eqref{eq:TAexcesswork} is nothing more than an average of the $W(t)$ taken in switching times of a uniform random variable in the range $[0,\tau]$. Therefore, the time-averaged work can be measured in the laboratory considering an average in the data set of processes executed in the following way: first, we choose a switching time $\tau$. After, we randomly choose an initial condition from the canonical ensemble and a time $t$ from a uniform distribution in the range $[0,\tau]$ to avoid preferred works in the time average. Removing the heat bath, we perform the works by changing the external parameter, collecting their values at the end. The data set produced will furnish, on average, the time-averaged work. I remark that I considered that this unique average will take care of the two random processes of this process: in the initial state and in the switching time. This procedure seems therefore to be faster than the conventional method of calculating the averaged works for equally spaced switching times and taking the average.

\section{Example: Landau-Zener model}
\label{sec:example}

To exemplify our results, we are going to treat here the Landau-Zener model at initial temperature $T=0$. The objective is to bring one more example to the literature that furnishes an oscillatory relaxation function. Systems that are characterized by this kind of feature represent an important class in the literature, ever since it includes systems whose validity in Hamiltonian modeling is out of the debate. It includes, for example, the classical and quantum harmonic oscillators \cite{acconcia2015a,acconcia2015b}, the 1/2 spin interacting with the magnetic field \cite{bonanca2016} and statistical anyons \cite{myers2021}.

The Hamiltonian of the system is
\be
\mc{H} = \Delta\sigma_x+\Gamma(t)\sigma_z,
\ee
where $\sigma_x$, $\sigma_z$ are the Pauli matrices in the $x$ and $z$ direction respectively, $\Delta$ the coupling energy and $\Gamma(t)=\Gamma_0+g(t)\delta\Gamma$ the time-dependent magnetic field.
Calculating the relaxation function, we have
\be
\Psi_0(t) = \mc{A}\co{\Omega t},
\ee
where
\be
\mc{A}=\frac{2\Delta^2}{\Delta^2+\Gamma_0^2}\frac{\Gamma_0+\sqrt{\Delta^2+\Gamma_0^2}}{\Delta^2+\left(\Gamma_0+\sqrt{\Delta^2+\Gamma_0^2}\right)^2},
\ee

\be
\Omega=\frac{2}{\hbar}\sqrt{\Delta^2+\Gamma_0^2}.
\ee

\subsection{Treatment with averaged work}

\begin{figure}
    \centering
    $(a)$\\
    \includegraphics[scale=0.5]{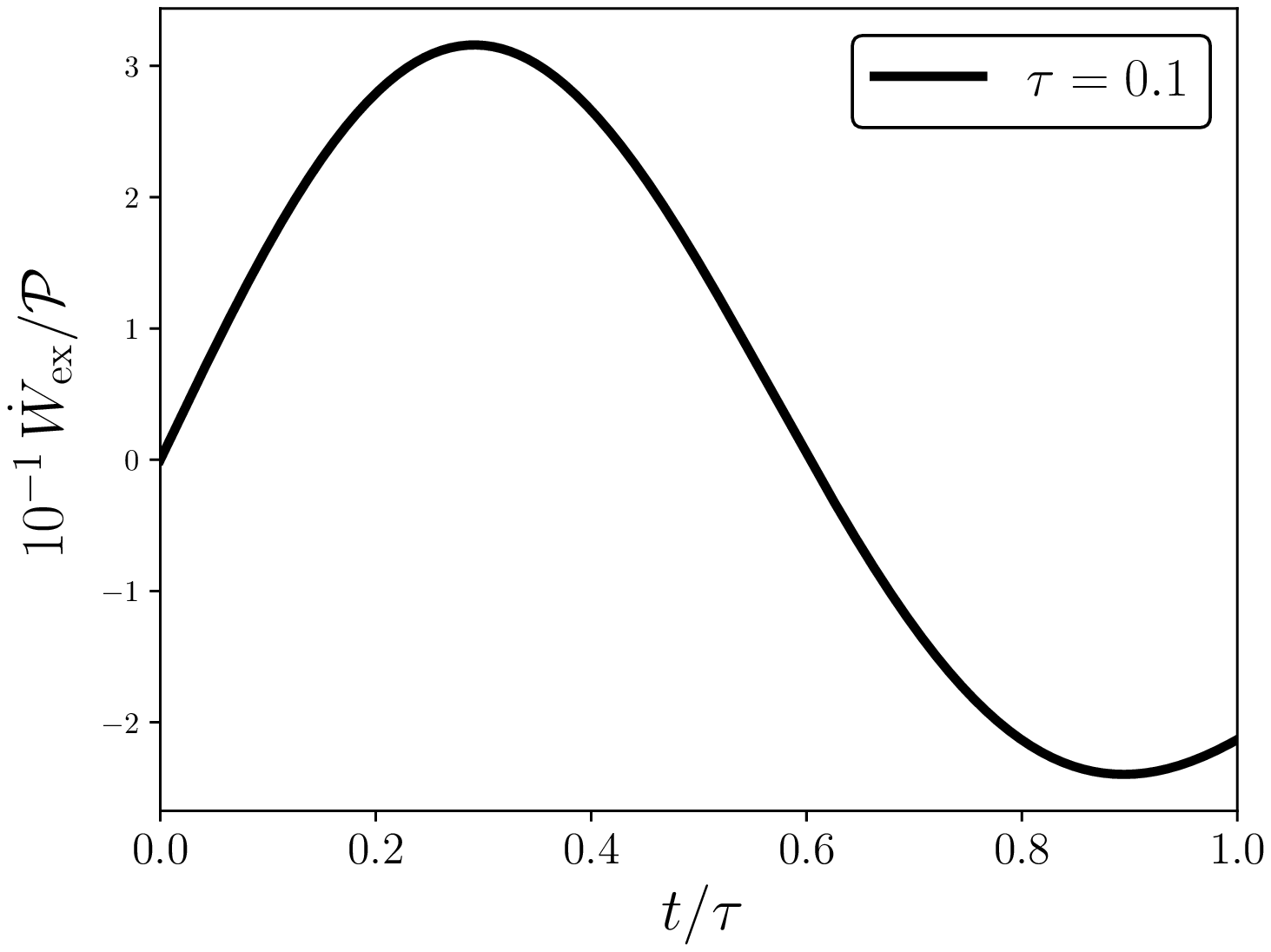}
    
    $(b)$\\
    \includegraphics[scale=0.5]{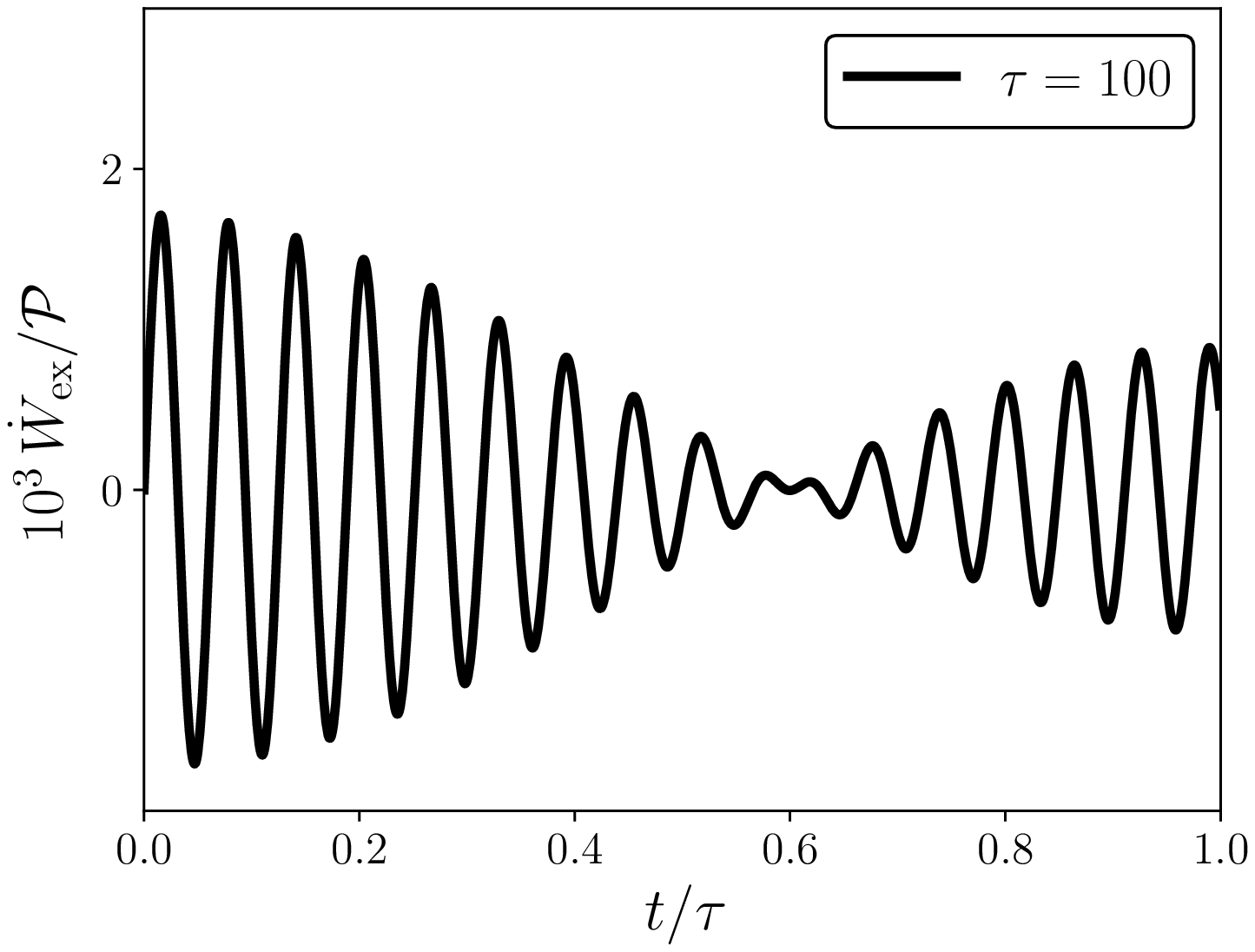}
    \caption{Entropy production rate for Landau-Zener model using Eq. \eqref{eq:protocol} as a protocol. Graphic $(a)$ depicts the entropy production rate for $\tau=0.1$, showing negative values in that quantity. Graphic $(b)$ depicts for $\tau=10$, showing as well negative values in the entropy production rate. The mechanism of associating the existence of a negative entropy production rate with the proximity of equilibrium does not hold anymore in this context. It was used $\Omega=1$.}
    \label{fig:3}
\end{figure}

First, we show that this function satisfies the 2nd Law of Thermodynamics \eqref{eq:2ndstate} using the criteria for linear-response theory. Second, we present the problems that are generated by using the traditional averaged excess work. Indeed, the Laplace transform and Fourier transform are
\be
\widetilde{\Psi}(s) = \frac{\mathcal{A}s}{s^2+\Omega^2}<\infty,\quad \hat{\Psi}(\omega) = \sqrt{\frac{\pi}{8}}(\delta(\omega)+2\delta(0))\ge 0. 
\ee
For the second point, the main problem with the average excess work treatment is the fact that the relaxation function does not decorrelate for long times, that is, there is no convergence to a Dirac delta in this regime. As a first consequence, the extension of the treatment of the finite-time and weak regimes to the slowly-varying ones by using Eq.~\eqref{eq:approximationSL} is lost. Also, the system presents a mathematically ill-defined relaxation time. Indeed
\be
\tau_R = \frac{\sin{(\infty)}}{\Omega}.
\label{eq:randomrelaxtime}
\ee
Observe that such relaxation time can be physically interpreted as a random number between $-1/\Omega$ and $1/\Omega$. Indeed, when measured in the laboratory using its own definition, it will furnish a random quantity since the sum of the integral must stop at some finite but high instant of time. Also, we speculate that such a randomness property is allowed for each instant of time. Indeed, as the relaxation function depends on $\Omega=\sin{(\infty)}/\tau_R$, at each time $t$, the sine could oscillate such that $\tau_R$ do it in the same fashion. Therefore, the relaxation time would be limited random number along the process. Finally, the mechanism found to justify the existence of a negative entropy production rate is not suited for this kind of system. Indeed, considering a driving process where the protocol is given by
\be
g(t)=\frac{t}{\tau}+\si{\frac{\pi t}{\tau}},
\label{eq:protocol}
\ee
with $\Omega=1$, we depict the entropy production rate for different $\tau$ in Fig.~\ref{fig:3}, observing that in all cases there are instants of time where its sign is negative.

\subsection{Compatibility with the Second Law}

We now analyze the Laplace and Fourier transforms of the time-averaged relaxation function to see if they agree with the compatibility criteria. The time-averaged relaxation function is
\be
\overline{\Psi}_0(t) = \mc{A}\,{\rm sinc}(\Omega t),
\label{eq:oscfunc}
\ee
with the respective time-averaged relaxation time
\be
\overline{\tau}_R=\frac{\pi}{2\Omega}.
\ee
Its Laplace and Fourier transforms are
\be
\widetilde{\overline{\Psi}}_0(s) = \frac{\mc{A}}{\Omega}\arctan{\left(\frac{\Omega}{s}\right)}<\infty
\ee
\be
\Hat{\overline{\Psi}}_0(\omega) =\frac{\sqrt{\pi}\mc{A}}{2\sqrt{2}\Omega}( {\rm sign}{(\omega+\Omega)}-{\rm sign}{(\omega-\Omega)})\ge 0,
\ee
therefore the Landau-Zener model, and any other system with an oscillatory relaxation function of type \eqref{eq:oscfunc}, agrees with the Second Law of Thermodynamics of Eq. \eqref{eq:2ndLawTimeAveragingExcessWork2}.

\subsection{Slowly-varying processes}

As a main consequence of having a relaxation time, the time-averaged excess work in the slowly-varying processes is given by
\be
\overline{W}_{\rm ex}(t) = \int_0^\tau\overline{\tau}_R[\Gamma(t)]\overline{\chi}[\Gamma(t)]\Gamma^2(t)dt,
\ee
where
\be
\overline{\tau}_R[\Gamma(t)]=\frac{\pi}{2\Omega[\Gamma(t)]},\quad \Omega[\Gamma(t)] = \frac{2}{\hbar}\sqrt{\Delta^2+\Gamma^2(t)},
\ee
and
\be
\overline{\chi}[\Gamma(t)] =\frac{2\Delta^2}{\Delta^2+\Gamma^2(t)}\frac{\Gamma(t)+\sqrt{\Delta^2+\Gamma^2(t)}}{\Delta^2+\left(\Gamma(t)+\sqrt{\Delta^2+\Gamma^2(t)}\right)^2}.
\ee

\subsection{Entropy production rates}

\begin{figure}
    \centering
    $(a)$\\
    \includegraphics[scale=0.5]{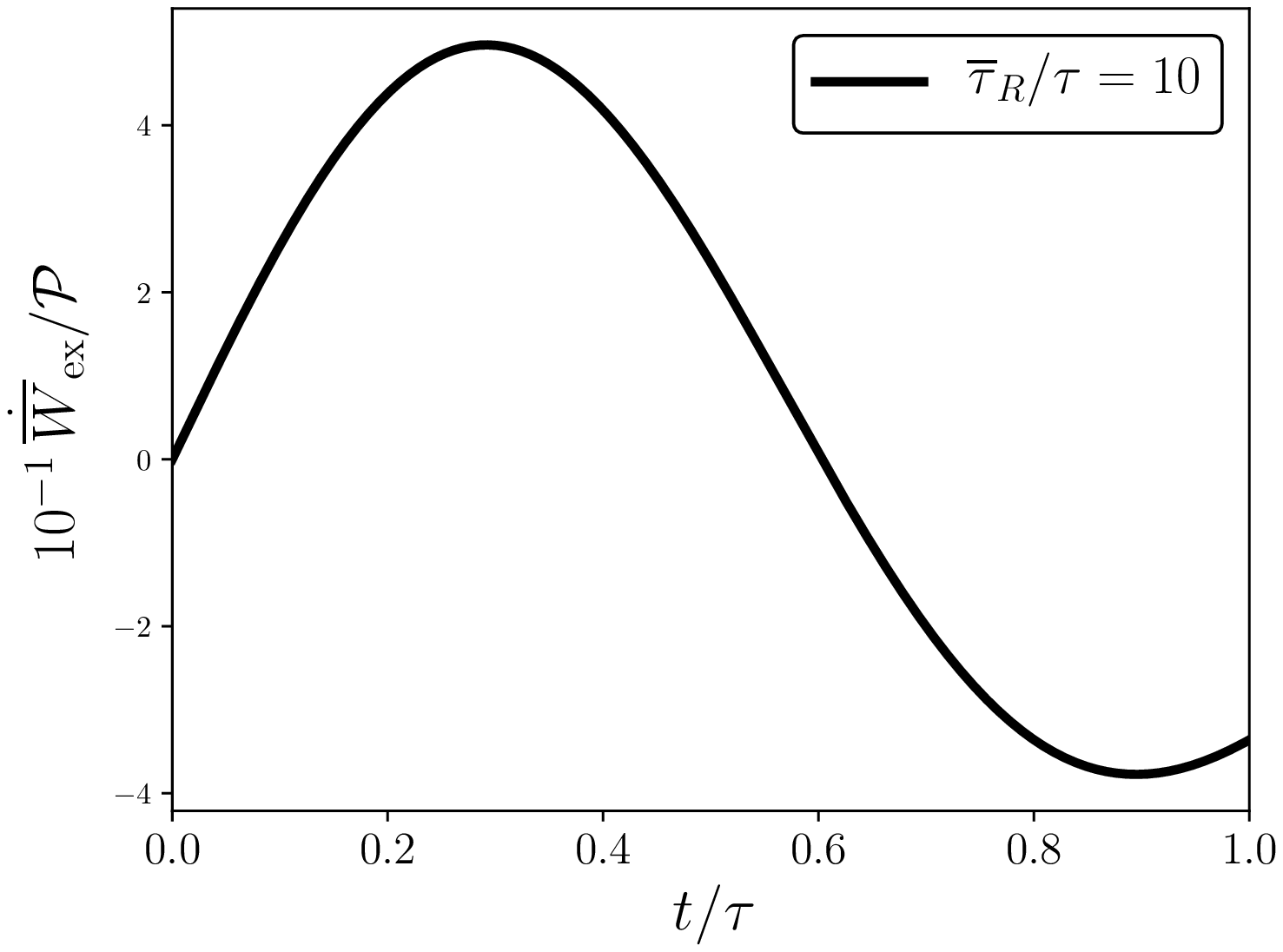}
    
    $(b)$\\
    \includegraphics[scale=0.5]{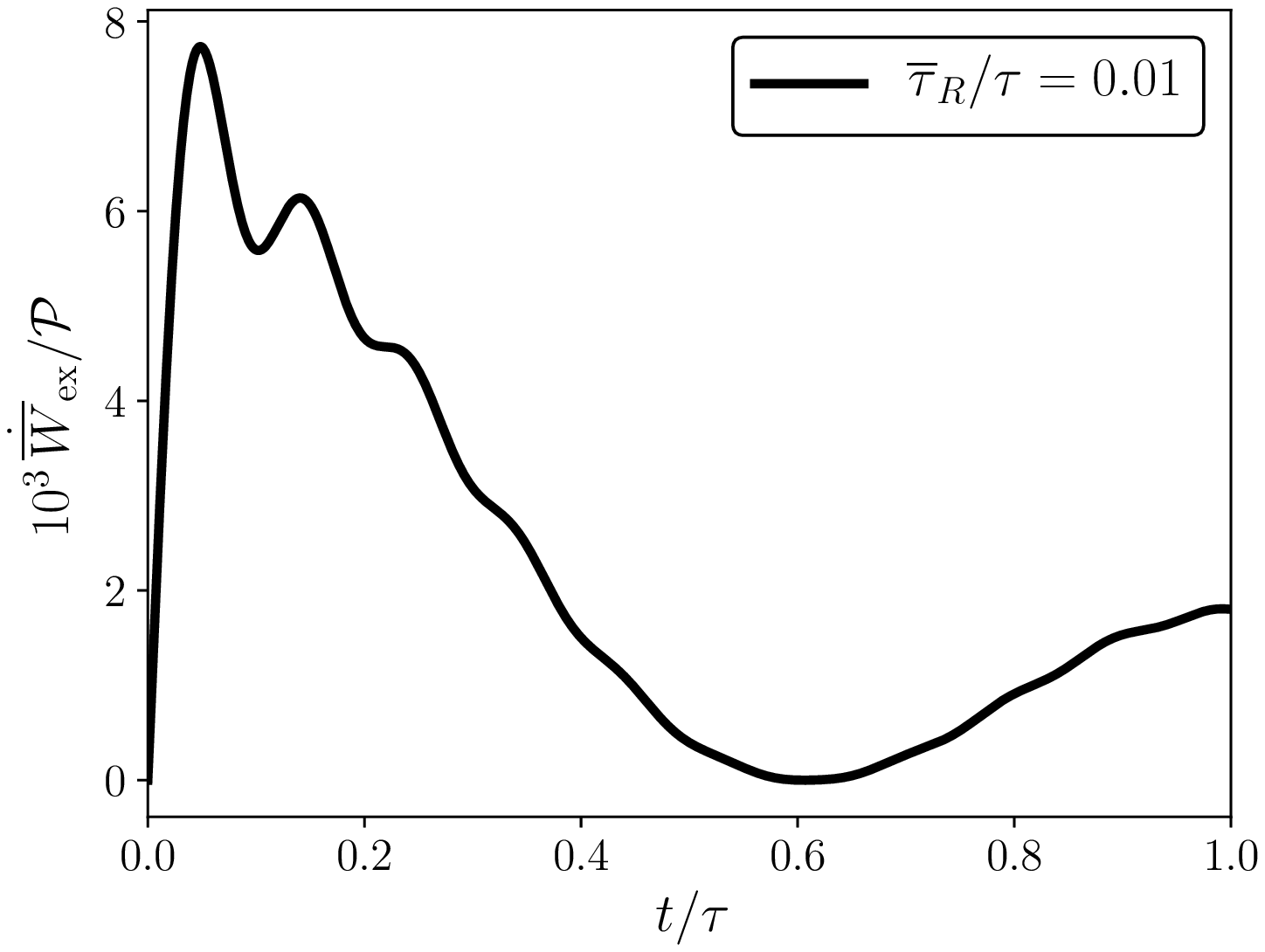}
    \caption{Entropy production rate for Landau-Zener model using Eq. \eqref{eq:protocol} as a protocol. Graphic $(a)$ depicts the entropy production rate for $\overline{\tau}_R/\tau=10$, showing negative values in that quantity. Graphic $(b)$ depicts for $\overline{\tau}_R/\tau=0.01$, showing no negative values in the entropy production rate. The behavior is identical to the case with isothermal processes.}
    \label{fig:2}
\end{figure}

Considering a driving process where the protocol is that of Eq.~\eqref{eq:protocol}, we depict the entropy production rate for different ratios $\overline{\tau}_R/\tau$ on Fig.~\ref{fig:2}. The situation is analogous to the case of isothermal processes \cite{naze2020}. For $\overline{\tau}_R/\tau\ll 1$, we observe only positive entropy production rate, while, for $\overline{\tau}_R/\tau\gg 1$, we observe negative values.

\section{Final remarks}
\label{sec:final} 

This work was divided into two parts. In the first, we identified the criteria that make compatible the linear-response theory with the Second Law of Thermodynamics for thermally isolated systems with bounded relaxation functions. We observe that they are the same as those for the case of isothermal processes. In the second part, observing the role of the time-averaged work in the identification of the previous criteria, we explored the possible consequences of defining it as the main quantity of interest instead of its traditional averaged counterpart. This proceeding defines an unexpected relaxation time to thermally isolated systems with oscillatory relaxation function, which allows them to obtain some mathematical properties of isothermal processes: the construction of slowly-varying processes from linear-response theory and to observe negative entropy production rates for non-monotonic protocols in rapid processes. We illustrate our results with the Landau-Zener model. Last but not least, the example treated illustrates how similar are thermally isolated and isothermal processes. Indeed, the time-averaged excess and irreversible works are of the same nature: both are composed of an average of the relaxation function, which has, each one at its nature, an intrinsic random relaxation time.  

\begin{acknowledgments}
The author acknowledges Marcus V. S. Bonan\c{c}a for suggestions in this paper.
\end{acknowledgments}

\appendix

\section{Proving equivalences of Eqs.~(18)-(20)}
\label{app:A}

\subsection{Proving Eq.~(18)}

First, we prove the direct implication
\be
\hat{\Psi}(0)<\infty \Rightarrow \lim_{\tau\rightarrow\infty}\frac{1}{\tau}\int_0^\tau \Psi(t)dt=0.
\ee
We have
\begin{align*}
     \lim_{\tau\rightarrow\infty}\frac{1}{\tau}\int_0^\tau \Psi(t)dt
    &= \lim_{\tau\rightarrow\infty}\frac{1}{\tau}\lim_{\tau\rightarrow\infty}\int_0^\tau \Psi(t)dt\\
    &= \lim_{\tau\rightarrow\infty}\frac{1}{\tau}\int_0^\infty \Psi(t)dt\\
    &=\lim_{\tau\rightarrow\infty}\frac{1}{\tau}\hat{\Psi}(0).
\end{align*}
Since $\Psi(t)$ is bounded, so $\hat{\Psi}(0)$ is as well. The reverse implication is a direct consequence of the final value theorem (Eq.~\eqref{eq:appfvt}).

\subsection{Proving Eq.~(19)}

The direct implication
\be
 \lim_{\tau\rightarrow\infty}\frac{1}{\tau}\int_0^\tau \Psi(t)dt=0\Rightarrow \lim_{\tau\rightarrow\infty}\frac{1}{\tau}\int_0^\tau W_{\rm ex}(t)dt=0
\ee
is proven considering the definition~\eqref{eq:TAexcesswork}. In the reverse implication, we still use Eq.~\eqref{eq:TAexcesswork} and consider Bochner's theorem.

\subsection{Proving Eq.~(20)}

In the direct implication
\begin{align}
\lim_{\tau\rightarrow \infty} \frac{1}{\tau}\int_0^\tau W_{\rm ex}(t)dt=0 &\Rightarrow \lim_{\tau\rightarrow \infty} W_{\rm ex}(\tau)=0,
\label{eq:equivasymplim}
\end{align}
we consider
\be
\lim_{\tau\rightarrow \infty} \frac{1}{\tau}\int_0^\tau W_{\rm ex}(t)dt=\int_0^1\left(\lim_{\tau\rightarrow\infty}W_{\rm ex}(\tau u)\right)du
\ee
and use the fact that $W_{\rm ex}(\tau)\ge 0, \forall \tau$. The reverse implication is direct.

\section{Time-averaged relaxation function}
\label{app:B}

Let us deduce the time-averaged relaxation function. Using the assumption that $\lambda(t)=\lambda(t/\tau)$, we have
\begin{align*}
    2\overline{W}_{\rm ex}(\tau)&=\frac{1}{\tau}\int_0^\tau\int_0^t\int_0^t\Psi(u-u')\dot{\lambda}(u)\dot{\lambda}(u')dudu'dt\\
    &=\frac{1}{\tau}\int_0^\tau\int_0^1\int_0^1\Psi(t(v-v'))\dot{\lambda}(v)\dot{\lambda}(v')dvdv'dt\\
    &=\int_0^1\int_0^1\left(\frac{1}{\tau}\int_0^\tau\Psi(t(v-v'))dt\right)\dot{\lambda}(v)\dot{\lambda}(v')dvdv'\\
    &=\int_0^1\int_0^1\left(\frac{1}{\tau(v-v')}\int_0^{\tau(v-v')}\Psi(y)dy\right)\dot{\lambda}(v)\dot{\lambda}(v')dvdv'\\
    &=\int_0^1\int_0^1\overline{\Psi}(\tau(v-v'))\dot{\lambda}(v)\dot{\lambda}(v')dvdv'\\
    &=\int_0^\tau\int_0^\tau\overline{\Psi}(u-u')\dot{\lambda}(u)\dot{\lambda}(u')dudu',
\end{align*}
where
\be
\overline{\Psi}(t)=\frac{1}{t}\int_0^t\Psi(u)du
\ee
is the time-averaged relaxation function. Observe that it preserves the symmetry of the relaxation function, that is, $\overline{\Psi}(t)=\overline{\Psi}(-t)$.

\bibliography{APIP.bib}
\bibliographystyle{apsrev4-2}

\end{document}